\documentclass{ws-mpla}
\usepackage[super]{cite}
\usepackage{graphicx}

\usepackage{mathrsfs,amsbsy,amssymb,latexsym,amsfonts,amsmath,amssymb,}

\newcommand\CI{\mathcal{I}}

\newcommand\CN{\mathcal{N}}

\newcommand\CQ{\mathcal{Q}}

\newcommand\bbC{\mathbb{C}}

\newcommand\nn{\nonumber}

\newcommand\pa{\partial}

\newcommand\adss[2]{AdS$_{#1}\times$S$^{#2}$}

\renewcommand\d{{\rm d}}

\newcommand\diag{{\rm diag}}

\newcommand\bref[1]{(\ref{#1})}

\begin{document}

\markboth{Fumiya Takeuchi and Makoto Sakaguchi}
{A Note on Supersymmetries in AdS$_5$/CFT$_4$}
\catchline{}{}{}{}{}

\title{A Note on Supersymmetries in AdS$_5$/CFT$_4$}

\author{\footnotesize Fumiya Takeuchi}

\address{Graduate School of Science and Engineering, Ibaraki University, Mito 310-8512, Japan\\
15nd404g@vc.ibaraki.ac.jp}

\author{Makoto Sakaguchi}

\address{Department of Physics, Ibaraki University, Mito 310-8512, Japan\\
makoto.sakaguchi.phys@vc.ibaraki.ac.jp}

\maketitle

\pub{Received (Day Month Year)}{Revised (Day Month Year)}

\begin{abstract}
The $\CN=4$ superconformal algebra is derived from
the symmetry transformations of fields in the $\CN=4$ SYM action in $D=4$.
We use a Majorana-Weyl spinor in $D=10$ instead of four Weyl spinors in $D=4$.
This makes it transparent to relate generators of the $\CN=4$ superconformal algebra
to those of  the super-\adss{5}{5} algebra.
Especially, we obtain the concrete map from the supersymmetries $Q$ and conformal supersymmetries $S$ in $\CN=4$ SYM to the supersymmetries $(\CQ_1,\CQ_2)$ in the \adss{5}{5} background.

\end{abstract}

\ccode{PACS Nos.: 11.25.Tq; 11.30.-j; 12.60.Jv}

\section{Introduction}	

The gauge/gravity duality\,\cite{gauge/gravity}
 continues to attract much interest
as it may provide hints for the non-perturbative formulation of M-theory and superstring theories.
The  AdS/CFT duality\,\cite{AdS/CFT,AdS/CFT2} 
has been investigated
 as a concrete realization of it.
A typical example
is the AdS$_5$/CFT$_4$ duality
which is a duality between 
the IIB superstring theory in the \adss{5}{5} background
and
the $\CN=4$ super Yang-Mills (SYM) theory
in $D=4$.
It is noted that 
both the super-isometry algebra of the \adss{5}{5} background 
and the superconformal algebra of the $\CN=4$ SYM
are psu(2,2$|$4).
This gives a fundamental non-trivial check for this duality.

It is known that the
$\CN=4$ SYM action in $D=4$
may be obtained as a dimensional reduction from 
the $\CN=(1,0)$ SYM action in $D=10$.
It follows that four Weyl spinors in $D=4$
can be represented by a $D=10$ Majorana-Weyl (MW) spinor. 
We use the $D=10$ MW spinor
throughout this paper.
One of the advantages of using the $D=10$ MW spinor notation is
to make it transparent to explore supersymmetries in  AdS$_5$/CFT$_4$.

In this note, we will derive the concrete transformation
laws of fields in the $\CN=4$ SYM under the PSU(2,2$|$4)
in  the $D=10$ MW spinor notation.
It is shown that 
the supersymmetry charge $Q$ is a MW spinor of positive chirality
while the conformal supersymmetry charge $S$ is that of negative chirality.
In the AdS$_5$/CFT$_4$ duality, 
these must be related to the supersymmetry charges
$(\CQ_1,\CQ_2)$
of positive chirality
on the \adss{5}{5} background.
The thrust of this note is to give a concrete map
from the generators of the $\CN=4$ superconformal algebra
to those of the super-\adss{5}{5} algebra,
including fermionic charges.

In the next section, we will begin with deriving transformation laws of 
the $\CN=(1,0)$ super-Poincar\'e symmetry
by examining  the SYM action in $D=10$.
It is shown in section 3 that they reduce  to
the $D=4$ Poincar\'e symmetry,
SU(4) R-symmetry
and 16 supersymmetries $Q$
of the SYM action in $D=4$.
Furthermore the SYM in $D=4$ is shown to acquire extra enhanced  symmetries:
special conformal symmetry, scale symmetry and 16 conformal supersymmetries $S$,
so that it has the $D=4$, $\CN=4$ superconformal symmetry in total.
In section 4, 
we will give a concrete map between the superconformal algebra in $D=4$ and
the supersymmetry algebra of \adss{5}{5}\,\cite{HKS02}.
Our map shows that
the conformal supersymmetries $S$  
are the supersymmetries
which are  broken in the presence of a D3-brane
and are restored in the near horizon limit,
as expected.
In the final section,
we comment on the central charges realized in terms of the $\CN=4$ SYM fields.

\section{$\CN=(1,0)$ super Yang-Mills theory in $D=10$}

We will show that the action of the $D=10$ SYM has 
the $\CN=(1,0)$ super-Poincar\'e
symmetry in $D=10$.
The  $\CN=(1,0)$ SYM action
we examine is given as 
\begin{align}
S=& \int \d^{10}x \left[
-\frac{1}{4} F_{MN}^a F^{aMN}
-\frac{i}{2}\bar \psi^a \Gamma^M D_M\psi^a
\right]~,
\label{action 10}
\end{align}
where we have defined the following objects
\begin{align}
F^{a}_{MN}&= \pa_M A^a_N-\pa_N A^a_M +g f^{abc}A_M^bA_N^c~,~~~
D_M\psi^a=\pa_M\psi^a+gf^{abc}A_M^b\psi^c~
\end{align}
and  $M,N=0,1,\cdots,9$ denote spacetime indices.
The fields $A_M=A^a_MT^a$ and $\psi=\psi^a T^a$ transform in the adjoint representation of
the gauge group U($N$)
where  $N\times N$ matrices $T^a$ ($a=1,2,\cdots,N^2$)  are the
generators of  U($N$)
satisfying $[T^a,T^b]=if^{abc}T^c$.
The Gamma-matrices $\Gamma^M$ in  ($1+9$)-dimensions are 32$\times$32 matrices satisfying
$\{\Gamma^M,\Gamma^N\}=2\eta^{MN}$
where $\eta^{MN}=\mathrm{diag}(-1,+1,\cdots,+1)$.
The $\psi^a$ are
MW spinors in  ($1+9$)-dimensions
which are 32 component column vectors satisfying $h_+\psi^a=\psi^a$,
where we have defined $h_\pm\equiv \frac{1}{2}(1\pm\Gamma_{11})$ 
with $\Gamma_{11}=\Gamma^{0123\cdots 9}$.
We define $\bar \psi^a\equiv \psi^a{}^T C$ where $C$ is the charge conjugation matrix
satisfying
$C\Gamma_MC^{-1}=-\Gamma_M^T$,
$C^\dag C=1$ and $C^T=-C$.

The action \bref{action 10} is invariant under the following
symmetries.\footnote{
The superscript on $\delta$ denotes the parameter of the corresponding transformation.}
\begin{itemlist}

\item{Poincar\'e symmetry}
generated by the translation $P_M$ with a parameter $a^M$
\begin{align}
\delta_P^a A_M^a&= a^N\pa_N A_M^a~,~~
\delta_P^a\psi^a= a^N\pa_N \psi^a~,
\end{align}
and the Lorentz rotation $L_{MN}$ with a parameter $\omega^{MN}=-\omega^{NM}$
\begin{align}
\delta_L^\omega A_M^a&=  x^P\omega_P{}^N \pa_NA_M^a+\omega_M{}^NA_N^a~,~~
\delta_L^\omega\psi^a= x^P\omega_P{}^N\pa_N\psi^a+\frac{1}{4}\omega^{MN}\Gamma_{MN}\psi^a~.
\end{align}

\item
{$\CN=(1,0)$ supersymmetry}
generated by the supertranslation $Q$
with a Grassmann-odd MW spinor parameter $\epsilon$
of positive chirality
$h_+ \epsilon = \epsilon$
\begin{align}
\delta_Q^\epsilon A_M^a&= \frac{i}{2} \bar \epsilon \Gamma_M\psi^a~,~~
\delta_Q^\epsilon \psi^a= -\frac{1}{4} F^a_{MN}\Gamma^{MN}\epsilon~.
\end{align}
A key relation for this symmetry is the Fierz identity
\begin{align}
(C\Gamma_M)_{\alpha(\beta}(C\Gamma^M)_{\gamma\delta)}=0~,
\label{10D Fierz}
\end{align}
which  is proven in Appendix A as  \bref{Fierz 1}.

\item
{Gauge symmetry}
generated by
a local gauge transformation with a parameter $\lambda^a(x)$
\begin{align}
\delta_g^\lambda A_M^a= D_M\lambda^a(x) = \pa_M \lambda^a +gf^{abc}A_M^b \lambda^c~,~~
\delta_g^\lambda \psi^a=gf^{abc}\psi^b \lambda^c~.
\end{align}
\end{itemlist}

\subsection{$\CN=(1,0)$ super-Poincar\'e symmetry in $D=10$}

We shall derive the symmetry algebra of the above symmetries.
For this purpose, we need to know  commutators among $\{\delta_L,\delta_P,\delta_Q\}$.
Commutators among $\{\delta_L,\delta_P\}$
are obtained as
\begin{align}
[\delta_P^a,\delta_P^{a'}]\Phi&=0~,~~~
\label{d PP 10}\\
{[}\delta_L^\omega,\delta_P^a]\Phi&=\delta_P^{\tilde a}\Phi
~~~\mbox{with}~~~
\tilde a^M=a^N\omega_{N}{}^M~,
\label{d LP 10}\\
{[}\delta_L^\omega,\delta_L^{\omega'}]\Phi&=\delta_L^{\tilde \omega}\Phi
~~~\mbox{with}~~~
\tilde\omega_M{}^N=\omega'_M{}^P\omega_P{}^N
-\omega_M{}^P\omega'_P{}^N~.
\label{d LL 10}
\end{align}
In this section, 
we will denote collectively the fields 
$A^a_M$ and $\psi^a$
simply as 
$\Phi$.
Next, we examine commutators between $\delta_Q$ and one of $\{\delta_L,\delta_P,\delta_Q\}$.
One finds
\begin{align}
[\delta_Q^\epsilon, \delta_Q^{\epsilon'}]\Phi&=\delta_P^{\tilde a} \Phi +\delta_g^\lambda \Phi~~~
\mbox{with}~~\tilde a^N=-\frac{i}{2}\bar \epsilon\Gamma^N \epsilon'
~~\mbox{and}~~
\lambda^a=\frac{i}{2} \bar\epsilon \Gamma^N\epsilon' A_N^a~,
\label{10D QQ psi}
\\
[\delta_Q^\epsilon,\delta_P^a]\Phi&=0~,~~~\\
[\delta_Q^\epsilon,\delta_L^\omega]\Phi &=\delta_Q^{\tilde\epsilon} \Phi~~~
\mbox{with}~~~
{\tilde\epsilon}=\frac{1}{4}\omega^{MN}\Gamma_{MN}\epsilon~.
\end{align}
In eqn (\ref{10D QQ psi}), we have used the equation of motion
\begin{align}
\Gamma^MD_M \psi^a=0~
\label{10D eom psi}
\end{align}
and the Fierz identity \bref{10D Fierz}.

We may define
generators $Q_\alpha$, $P_M$ and $L_{MN}$
by
\begin{align}
\delta_Q^\epsilon\Phi=-i[ Q_\alpha\epsilon^\alpha,\Phi] ~,~~~
\delta_P^a\Phi=-i[a^N P_N,\Phi]~,~~~
\delta_L^\omega\Phi=-i[\frac{1}{2}w^{MN}L_{MN},\Phi]~,
\end{align}
where $Q$ is a $D=10$ MW spinor of positive chirality $Q h_+=Q$,
and a 32-component row vector.
On the right hand side of $[\delta_Q^\epsilon, \delta_Q^\eta] $, 
there appears a gauge transformation $\delta_g$.
This is because $A_M^a$ and $\psi^a$
are not gauge invariant.
So we may omit  $\delta_g\Phi$ in considering spacetime symmetry algebra.
Let us consider the equation $[\delta_Q^\epsilon, \delta_Q^{\epsilon'}]\Phi=\delta_P^{\tilde a} \Phi$.
The left hand side means
\begin{align}
[\delta_Q^\epsilon,\,\delta^{\epsilon'}_Q]\Phi
=[Q\epsilon,[Q\epsilon',\Phi]]
-[Q\epsilon',[Q\epsilon,\Phi]]
=[ [Q\epsilon,Q\epsilon'],\Phi]
~,
\end{align}
while the right hand side means $\delta^{\tilde a}_P \Phi =-i[\tilde a^NP_N, \Phi]$.
It follows from these expressions that
\begin{align}
[Q\epsilon,Q\epsilon']=-i\tilde a ^NP_N.
\end{align}
Furthermore, since 
\begin{align}
[Q\epsilon,Q\epsilon']&=-\epsilon^\alpha\{Q_\alpha,Q_\beta\}\epsilon'^\beta~,\nn\\
-i\tilde a^NP_N&=-i\left( -\frac{i}{2}\bar\epsilon \Gamma^N\epsilon' \right) P_N
=-\epsilon^\alpha\frac{1}{2}(C\Gamma^N)_{\alpha\beta} P_N \epsilon'^\beta~,
\end{align}
we obtain
\begin{align}
\{Q_\alpha,\,Q_\beta\}=\frac{1}{2}(C\Gamma^Nh_+)_{\alpha\beta} P_N~.
\label{QQ 10}
\end{align}
Similarly,
$[\delta_Q^\epsilon,\delta_L^\omega]\Phi=\delta_Q^{\tilde\epsilon}\Phi$,
${[}\delta_L^\omega,\delta_P^a]\Phi=\delta_P^{\tilde a}\Phi$
and
 ${[}\delta_L^\omega,\delta_L^{\omega'}]\Phi=\delta_L^{\tilde \omega}\Phi$
lead to
\begin{align}
[Q_\alpha,\,L_{MN}]&=-\frac{i}{2}(Q\Gamma_{MN})_\alpha
~,~~~
[L_{MN},P_P]= i(\eta_{NP}P_M -\eta_{MP}P_N)~,
\nn\\
[L_{MN},\,L_{PQ}]&=
i(\eta_{NP}L_{MQ}
-\eta_{MP}L_{NQ}
-\eta_{NQ}L_{MP}
+\eta_{MQ}L_{NP})~,
\label{bb 10}
\end{align}
respectively.
Summarizing,
we have shown that  the action \bref{action 10}
is invariant under the   $D=10$,  $\CN=(1,0)$   super-Poincar\'e symmetry
\bref{QQ 10}
and
\bref{bb 10}.

\section{$\CN=4$ super Yang-Mills theory in $D=4$}

\subsection{$D=4$ SYM from $D=10$ SYM}

We consider a dimensional reduction to four-dimensions by assuming that
\begin{align}
M=(\mu,i)~,~~
A_M^a=(A^a_\mu(x^\mu),\,\phi^a_i(x^\mu))~,~~
\psi^a=\psi^a(x^\mu)
\end{align}
where $\mu=0,1,2,3$ and $i=4,5,\dots,9$.
It is straightforward to see that the action \bref{action 10}
reduces to
\begin{align}
S=\int \d^4x &\left[
-\frac{1}{4}F^a_{\mu\nu}F^{a\mu\nu}
-\frac{1}{2}D_\mu \phi^{a}_iD^\mu\phi^a_i
-\frac{1}{4}g^2f^{abc}\phi^b_i\phi^c_j f^{ade}\phi^d_i\phi^e_j
\right.
\nonumber \\
&
\left.
-\frac{i}{2}\bar\psi^a\Gamma^\mu D_\mu\psi^a
-\frac{i}{2}\bar\psi^a\Gamma^i gf^{abc}\phi^b_i\psi^c
\right]
\label{SYM 4}
\end{align}
where  $F^{a}_{\mu\nu}$, $D_\mu\phi_i^a$ and $D_\mu\psi^a$
are defined by
\begin{align}
F^{a}_{\mu\nu}&= \pa_\mu A^a_\nu-\pa_\nu A^a_\mu +g f^{abc}A_\mu^bA_\nu^c~,~~~
D_\mu\phi_i^a=
\pa_\mu\phi_i^a +g f^{abc}A_\mu^b \phi_i^c
~,\nn\\
D_\mu\psi^a&=\pa_\mu\psi^a+gf^{abc}A_\mu^b\psi^c~.
\end{align}

\subsection{Symmetries inherited from $D=10$ SYM}

The following symmetries are inherited from  $D=10$ SYM.
It is straightforward to see that
the action \bref{SYM 4} is invariant  under these symmetry transformations.
Here and hereafter we will denote collectively 
$A_\mu^a,~\phi_i^a,$ and $\psi^a$ simply as $\Phi$.
\begin{itemlist}

\item 
{Translation symmetry $P_\mu$} with a parameter $a^\mu$
\begin{align}
\delta_P^a  \Phi=a^\nu\pa_\nu \Phi~.~~
\end{align}

\item 
{Lorentz symmetry $L_{\mu\nu}$} with a parameter $\omega^{\mu\nu}=-\omega^{\nu\mu}$
\begin{align}
\delta_L^\omega  A_\mu^a&= x^\sigma\omega_\sigma{}^\rho \pa_\rho A_\mu^a
+\omega_\mu{}^\nu A_\nu^a~,~~
\delta_L^\omega  \phi_i^a=x^\sigma\omega_\sigma{}^\rho \pa_\rho\phi_i ~,\nn\\
\delta_L^\omega \psi^a&=x^\sigma\omega_\sigma{}^\rho \pa_\rho \psi^a +\frac{1}{4}\omega^{\mu\nu}\Gamma_{\mu\nu}\psi^a~.
\end{align}

\item{SO(6)  symmetry $R_{ij}$}
with a parameter $m^{ij}=-m^{ji}$
\begin{align}
\delta_R^m  A_\mu^a&=0
~,~~~
\delta_R^m   \phi_i^a=m_i{}^j \phi_j^a ~,~~~
\delta_R^m  \psi^a=
\frac{1}{4}m^{ij}\Gamma_{ij}\psi^a~.
\end{align}

\item
{Supersymmetry $Q_\alpha$}
with a Grassmann-odd MW spinor parameter $\epsilon$
\begin{align}
\delta_Q^\epsilon  A^a_\mu&=
\frac{i}{2}\bar \epsilon \Gamma_\mu \psi^a~,~~~
\delta_Q^\epsilon \phi^a_i = \frac{i}{2}\bar \epsilon \Gamma_i \psi^a~,\nn\\
\delta_Q^\epsilon\psi^a&=
-\frac{1}{4}F^a_{\mu\nu}\Gamma^{\mu\nu} \epsilon
-\frac{1}{2} D_\mu\phi^a_i\Gamma^{\mu i}\epsilon
-\frac{1}{4} gf^{abc} \phi^b_i \phi^c_j \Gamma^{ij}\epsilon~.
\end{align}
The supersymmetry parameter $\epsilon$ has 16 nontrivial components,
while a Weyl spinor in four-dimensions
consists of two complex components. 
It follows that $\epsilon$ consists of four Weyl spinors in four-dimensions
so that the action is $\CN=4$ supersymmetric.

\item
{Gauge symmetry} with a parameter $\lambda^a(x)$
\begin{align}
\delta_g^\lambda A_\mu^a=
D_\mu\lambda^a= \pa_\mu\lambda^a +gf^{abc}A_\mu^b\lambda^c~,~~~
\delta_g^\lambda \phi_i^a=
gf^{abc}\phi_i^b\lambda^c~,~~~
\delta_g^\lambda \psi^a=
gf^{abc}\psi^b\lambda^c~.
\end{align}
\end{itemlist}

The commutators among $\{\delta_P,\delta_L,\delta_R\}$ are found to be
\begin{align}
[\delta_P,\delta_P]\Phi&=0~,~~~\\
{[}\delta_L^\omega,\delta_P^a]\Phi&=\delta_P^{\tilde a}\Phi
~~~\mbox{with}~~~
\tilde a^\mu=a^\nu\omega_\nu{}^\mu~,\\
{[}\delta_L^\omega,\delta_L^{\omega'}]\Phi&=\delta_L^{\tilde \omega}\Phi
~~~\mbox{with}~~~
\tilde\omega_\mu{}^\nu=\omega'_\mu{}^\rho\omega_\rho{}^\nu
-\omega_\mu{}^\rho\omega'_\rho{}^\nu~,\\
[\delta_R^m,\delta_R^{m'}]\Phi&=\delta_R^{\tilde m}\Phi
~~~\mbox{with}~~~
\tilde m_i{}^j=m'_i{}^km_k{}^j-m_i{}^km'_k{}^j~,
\end{align}
while those including $\delta_Q$ are found as
\begin{align}
[\delta_Q,\delta_P]\Phi&=0~,~~~\\
{[}\delta_Q^\epsilon,\delta_L^\omega]\Phi&=\delta_Q^{\tilde\epsilon}\Phi
~~~\mbox{with}~~~
\tilde\epsilon=\frac{1}{4}\omega^{\mu\nu}\Gamma_{\mu\nu}\epsilon~,\\
{[}\delta_Q^\epsilon,\delta_R^m]\Phi&=\delta_Q^{\tilde\epsilon}\Phi
~~~\mbox{with}~~~
\tilde\epsilon=\frac{1}{4}m^{ij}\Gamma_{ij}\epsilon~,\\
{[}\delta_Q^\epsilon,\delta_Q^{\epsilon'}]\Phi&=\delta_P^{\tilde a}\Phi+\delta_g^{\lambda}\Phi
~~~\mbox{with}~~~
\tilde a^\mu=\frac{i}{2}\bar\epsilon'\Gamma^\mu\epsilon
~~\mbox{and}~~
\lambda^a=-\frac{i}{2}A^a_\mu\bar\epsilon'\Gamma^\mu\epsilon
-\frac{i}{2}\phi_i^a\bar\epsilon'\Gamma^i\epsilon~.
\label{QQ psi}
\end{align}
In order to evaluate ${[}\delta_Q^\epsilon,\delta_Q^{\epsilon'}]\psi$ in  \bref{QQ psi},
we have used the equation of motion
\begin{align}
&\Gamma^\mu D_\mu\psi^a+gf^{abc}\phi^b_i\Gamma^i\psi^c=0~
\label{eom psi}
\end{align}
and the Fierz identity \bref{10D Fierz}.
As will be seen in section 3.4, 
they represent the $\CN=4$ super-Poincar\'e symmetry.

\subsection{Enhanced symmetries}
In addition to  the $\CN=4$ super-Poincar\'e symmetry
inherited from the  $D=10$ SYM, 
the $D=4$ SYM action
\bref{SYM 4} 
acquires the following enhanced symmetries:
the special conformal symmetry $K_\mu$,
the scale symmetry $D$
and 
the conformal supersymmetry $S_\alpha$.

\subsubsection*{Special conformal  symmetry $K_\mu$}
One can show  that the $D= 4$ SYM action \bref{SYM 4} 
is invariant under the special conformal transformation 
$K_\mu$
with a parameter $b^\mu$
\begin{align}
\delta_K^b A_\mu^a&= \delta x^\nu\pa_\nu A_\mu^a
-2(bx)A_\mu^a
+2(x_\mu b^\nu-b_\mu x^\nu)A_\nu^a~,\\
\delta_K^b\phi^a_i&=\delta x^\nu \pa_\nu \phi^a_i
-2(bx) \phi^a_i~,\\
\delta_K^b\psi^a&=
\delta x^\nu\pa_\nu \psi^a
-3
(bx)\psi^a
+\frac{1}{2}(x^\mu b^\nu - x^\nu b^\mu)\Gamma_{\mu\nu}\psi^a~,
\end{align}
where $\delta x^\nu=-2(bx)x^\nu
+b^\nu x^2$ and  $(bx)=b^\mu x_\mu$.
Note that the second terms in the right hand sides denote $-2\Delta(\Phi) (bx) \Phi$
where $\Delta(\Phi)$ stands for the dimension of $\Phi$,
i.e. $\Delta(A_\mu^a)=\Delta(\phi^a_i)=1$ and $\Delta(\psi^a)=3/2$.

Commutators between $\delta_K^b$
and one of $\{\delta_L,~
\delta_R,~\delta_K\}$
do not generate new transformations
\begin{align}
[\delta_K^b,\delta_L^\omega] \Phi&=\delta_K^{\tilde b}\Phi
~~~\mbox{with}~~~
\tilde b^\mu=-b^\nu\omega_\nu{}^\mu~,\\
{[}\delta_K,\delta_K]\Phi&=0~,\\
{[}\delta_K,\delta_R]\Phi&=0
~,
\end{align}
but $[\delta_K,\delta_P]\Phi$ and $[\delta_K,\delta_Q]\Phi$
are shown to require extra transformations $\delta_D$ and $\delta_S$.

\subsubsection*{Scale symmetry $D$}
The commutator between $\delta_K$ and $\delta_P$
is
\begin{align}
{[}\delta_K^b,\delta_P^a]\Phi&=\delta_L ^{\tilde \omega}\Phi 
-2a^\mu b_\mu
\left(x^\nu\pa_\nu \Phi +\Delta(\Phi)\Phi\right)
~~~\mbox{with}~~~
\tilde\omega^{\mu\nu}=2(a^\mu b^\nu-b^\mu a^\nu)~.
\label{KP D}
\end{align}
From \bref{KP D} we define
the scale transformation $D$ with a parameter $\alpha$ 
\begin{align}
\delta_D^\alpha \Phi&=
\alpha\left(x^\nu\pa_\nu \Phi +\Delta(\Phi)\Phi\right)~,
\end{align}
so that \bref{KP D} turns to ${[}\delta_K^b,\delta_P^a]\Phi=\delta_L ^{\tilde \omega}\Phi 
+\delta_D^{\tilde \alpha}\Phi$ with $\tilde\alpha=-2a^\mu b_\mu$.

\subsubsection*{Conformal supersymmetry $S_\alpha$}

We find that $[\delta_K^b,\delta_Q^\epsilon] \Phi$
takes the form
\begin{align}
[\delta_K^b,\delta_Q^\epsilon]A_\mu^a=&-\frac{i}{2}\bar\epsilon (b\Gamma) (x\Gamma)\Gamma_\mu \psi^a~,\\
{[}\delta_K^b,\delta_Q^\epsilon] \phi^a_i=&
-\frac{i}{2}\bar\epsilon (b\Gamma)  (x\Gamma)\Gamma_i \psi^a~,\\
{[}\delta_K^b,\delta_Q^\epsilon] \psi^a=&
\frac{1}{4}F_{\mu\nu}\Gamma^{\mu\nu}(x\Gamma)(b\Gamma)\epsilon
+\frac{1}{2}D_\mu\phi_i^a \Gamma^{\mu i}(x\Gamma)(b\Gamma)\epsilon
\nonumber\\&
+\frac{1}{4}gf^{abc}\phi^b_i\phi^c_j \Gamma^{ij}(x\Gamma)(b\Gamma)\epsilon
-\phi_i^a \Gamma^i(b\Gamma)\epsilon~.
\end{align}
From these expressions,
we define the conformal supersymmetry transformation $S_\alpha$
with a parameter $\eta$
\begin{align}
\delta_S^\eta A_\mu^a&=
\frac{i}{2}\bar\eta (x\Gamma)\Gamma_\mu \psi^a~,\\
\delta_S^\eta \phi^a_i&=
\frac{i}{2}\bar\eta (x\Gamma)\Gamma_i \psi^a~,\\
\delta_S^\eta \psi^a&=
\frac{1}{4}F_{\mu\nu}^a\Gamma^{\mu\nu}(x\Gamma)\eta
+\frac{1}{2}D_\mu\phi_i^a \Gamma^{\mu i}(x\Gamma)\eta
+\frac{1}{4}gf^{abc}\phi^b_i\phi^c_j \Gamma^{ij}(x\Gamma)\eta
-\phi_i^a \Gamma^i\eta~,
\end{align}
so that $[\delta_K^b, \delta_Q^\epsilon]\Phi=\delta_S^{\tilde\eta} \Phi~$
with $\tilde\eta=(b\Gamma)\epsilon$.
Note that $\eta$ has the opposite chirality to $\epsilon$ and $\psi$.

Nontrivial commutators including $\delta_D$ are
\begin{align}
[\delta_D^\alpha,\delta_P^a]\Phi&=\delta_P^{\tilde a}\Phi
~~~\mbox{with}~~~
\tilde a^\mu=\alpha a^\mu~,\\
{[}\delta_D^\alpha,\delta_K^b]\Phi&=\delta_K^{\tilde b}\Phi
~~~\mbox{with}~~~
\tilde b^\mu=-\alpha b^\mu~,\\
{[}\delta_D^\alpha,\delta_Q^\epsilon]\Phi&=\delta_Q^{\tilde \epsilon}\Phi
~~~\mbox{with}~~~
\tilde\epsilon=\frac{1}{2}\alpha\epsilon~,\\
{[}\delta_D^\alpha,\delta_S^\eta]\Phi&=\delta_S^{\tilde \eta}\Phi
~~~\mbox{with}~~~
\tilde \eta=-\frac{1}{2}\alpha\eta~.
\end{align}
Commutators including $\delta_S^\eta$ are found to be
\begin{align}
[\delta_S^\eta,\delta_L^\omega]\Phi&=\delta_S^{\tilde \eta}\Phi
~~~\mbox{with}~~~
\tilde \eta=\frac{1}{4} \omega^{\mu\nu}\Gamma_{\mu\nu}\eta~,\\
{[}\delta_S^\eta,\delta_R^m]\Phi&=\delta_S^{\tilde \eta}\Phi
~~~\mbox{with}~~~
\tilde \eta=\frac{1}{4}m^{ij}\Gamma_{ij}\eta~,\\
{[}\delta_P^a,\delta_S^\eta]\Phi&=\delta_Q^{\tilde \epsilon}\Phi
~~~\mbox{with}~~~
\tilde \epsilon=a^\mu\Gamma_{\mu}\eta~,\\
{[}\delta_S^\eta,\delta_S^{\eta'}]\Phi&=\delta_K^{\tilde b}\Phi+\delta_g^{\lambda}\Phi
\nn\\&
\mbox{with}~~~
\tilde b^\mu= -\frac{i}{2}\bar\eta \Gamma^\mu\eta'~~\mbox{and}~~
\lambda^a=-\frac{i}{2}\bar\eta(x\Gamma)(\phi^a_i \Gamma^i+A^a_\mu\Gamma^\mu)(x\Gamma)\eta'
~,
\label{SS psi}
\end{align}
and
\begin{align}
[\delta_Q^\epsilon,\delta_S^\eta]\phi^a_i&=
\delta_L^\omega \phi^a_i
+\delta_D^\alpha \phi^a_i
+\delta_R^m \phi^a_i
+\delta_g^\lambda \phi^a_i
~,~~\\
{[}\delta_Q^\epsilon,\delta_S^\eta]A^a_\mu&=
\delta_L^\omega A^a_\mu
+\delta_D^\alpha A^a_\mu
+\delta_g^\lambda A^a_\mu~,~~\\
{[}\delta_Q^\epsilon,\delta_S^\eta]\psi^a&=
\delta_L^\omega \psi^a
+\delta_D^\alpha \psi^a
+\delta_R^m \psi^a
+\delta_g^\lambda \psi^a~,
\label{[Q,S]psi}
\end{align}
where 
\begin{align}&
\omega^{\mu\nu}=-\frac{i}{2}\bar\epsilon \Gamma^{\mu\nu}\eta~,~~
\alpha=\frac{i}{2} \bar\epsilon \eta~,~~
m^{ij}=\frac{i}{2}\bar \epsilon \Gamma^{ij}\eta~,~~
\lambda=
-\frac{i}{2} \bar\epsilon (\phi^a_i\Gamma^i + A_\mu^a \Gamma^\mu) (x\Gamma) \eta
~.
\end{align}
In order to evaluate ${[}\delta_S^\eta,\delta_S^{\eta'}]\psi$ in  \bref{SS psi},
we have used the equation of motion \bref{eom psi}
and the Fierz identity \bref{10D Fierz}.
To show the equation \bref{[Q,S]psi}, 
we have used the following Fierz identity
\begin{align}
-i\epsilon\,(\bar\eta\psi)
-\frac{i}{2}\Gamma^M \eta\,(\bar\psi \Gamma_M \epsilon)=
-\frac{i}{4} \psi\,(\bar\epsilon \eta)
+\frac{i}{8} \Gamma^{MN}\psi\,(\bar\epsilon \Gamma_{MN}\eta)~,
\label{Fierz SS}
\end{align}
in addition to \bref{eom psi} and  \bref{10D Fierz}.
This is proven in Appendix A.

\subsection{$\CN=4$ superconformal algebra in $D=4$}

We may define
generators
$Q_\alpha,~P_\mu,~L_{\mu\nu},~R_{ij},~D,~K_\mu$ and $S_\alpha$
by
\begin{align}
\delta_Q^\epsilon\Phi&=-i[Q_\alpha\epsilon^\alpha,\Phi]~,~~
\delta_P^a\Phi=-i[a^\mu P_\mu,\Phi]~,~~
\delta_L^\omega\Phi=-i[\frac{1}{2}\omega^{\mu\nu}L_{\mu\nu},\Phi]~,~~
\nonumber\\
\delta_R^m\Phi&=-i[\frac{1}{2}m^{ij}R_{ij},\Phi]~,~~
\delta_D^\alpha\Phi=-i[\alpha D,\Phi]~,~~
\delta_K^b\Phi=-i[b^\mu K_\mu,\Phi]~,~~
\nonumber\\
\delta_S^\eta\Phi&=-i[S_\alpha\eta^\alpha,\Phi]~.
\end{align}
Following the method explained in section 2.1, 
we obtain the following commutation relations
\begin{align}
[L_{\mu\nu},P_\rho]&=
i(\eta_{\nu\rho}P_\mu-\eta_{\mu\rho}P_\nu)~,~
\nn\\
[L_{\mu\nu},L_{\rho\sigma}]&=
i(
\eta_{\nu\rho}L_{\mu\sigma}
-\eta_{\mu\rho}L_{\nu\sigma}
-\eta_{\nu\sigma}L_{\mu\rho}
+\eta_{\mu\sigma}L_{\nu\rho}
)~,
\nonumber\\
[R_{ij},R_{kl}]&=
i(\delta_{jk}R_{il}
-\delta_{ik}R_{jl}
-\delta_{jl}R_{ik}
+\delta_{il}R_{jk}
)~,\nn\\
[D,P_\mu]&=-iP_\mu~,~~
[D,K_\mu]=+iK_\mu~,~~
[L_{\mu\nu},K_\rho]=
i(\eta_{\nu\rho}K_\mu
-\eta_{\mu\rho}K_\nu)~,
\nn\\
[K_\mu,P_\nu]&=
2iL_{\mu\nu}
+2i \eta_{\mu\nu}D~,~~
\label{conf4}
\end{align}
and
\begin{align}
[Q_\alpha,L_{\mu\nu}]&=-\frac{i}{2}(Q\Gamma_{\mu\nu})_\alpha~,~~~
[Q_\alpha,R_{ij}]=-\frac{i}{2}(Q\Gamma_{ij})_\alpha~,~~~
[D,Q_\alpha]=-\frac{i}{2}Q_\alpha~,~~~
\nonumber\\
[S_\alpha,L_{\mu\nu}]&=-\frac{i}{2}(S\Gamma_{\mu\nu})_\alpha~,~~~
[S_\alpha,R_{ij}]=-\frac{i}{2}(S\Gamma_{ij})_\alpha~,~~~
[D,S_\alpha]=+\frac{i}{2}S_\alpha~,~~~
\nonumber\\
[K_\mu,Q_\alpha]&=-i(S\Gamma_{\mu})_\alpha~,~~~
[P_\mu,S_\alpha]=-i(Q\Gamma_{\mu})_\alpha~,~~~
\nonumber\\
\{Q_\alpha,Q_\beta\}&=\frac{1}{2}(C\Gamma^\mu h_+)_{\alpha\beta} P_\mu~,~~~
\{S_\alpha,S_\beta\}=\frac{1}{2}(C\Gamma^\mu h_-)_{\alpha\beta} K_\mu~,~~~
\nonumber\\
\{Q_\alpha,S_\beta\}&=\frac{1}{4}(C\Gamma^{\mu\nu} h_-)_{\alpha\beta} L_{\mu\nu}
-\frac{1}{2}(C h_-)_{\alpha\beta} D
-\frac{1}{4}(C\Gamma^{ij} h_-)_{\alpha\beta} R_{ij}~.
\label{super conf4}
\end{align}
It implies that the $\CN=4$ SYM in $D=4$ is invariant under the
$\CN=4$ superconformal transformations in $D=4$,
so that  it's symmetry algebra is psu(2,2$|4$).
The set of generators $\{P_\mu,~L_{\mu\nu},~R_{ij},~Q_\alpha\}$
forms a subalgebra of psu(2,2$|$4),
the $\CN=4$ super-Poincar\'e algebra.

\section{Relation to supersymmetries in AdS$_5\times$S$^5$ background}

In this section, we shall rewrite the $\CN=4$ superconformal algebra in $D=4$ given in
\bref{conf4} and \bref{super conf4}
into the supersymmetry algebra of the  \adss{5}{5} background.
This gives a concrete map between generators, including fermionic ones.

First, we examine bosonic generators
and show that they form {so(2,4)$\times$so(6)} of the isometry algebra of \adss{5}{5}.
Let us define $J_{MN}$ ($M,N=0,1,2,3,4,\sharp$) by
\begin{align}
J_{\mu\nu}=L_{\mu\nu}~,~~
J_{4\sharp}=-D~,~~
J_{\mu 4}=\frac{1}{2}(K_\mu-P_\mu)~,~~
J_{\mu \sharp}=\frac{1}{2}(K_\mu+P_\mu)~,
\end{align}
and then $J_{MN}$ are shown to satisfy so(2,4)
\begin{align}
[J_{MN},J_{PQ}]=
i(
\eta_{NP}J_{MQ}
-\eta_{MP}J_{NQ}
-\eta_{NQ}J_{MP}
+\eta_{MQ}J_{NP}
)
\end{align}
where
$\eta_{MN}=\diag(-,+,+,+,+,-)$
with $\eta_{00}=\eta_{\sharp\sharp}=-1$.
If we define $\hat J_{ab}$ and $\hat P_a$ ($a,b=0,1,3,4$) by
\begin{align}
\hat J_{ab}=J_{ab}~,~~~
\hat P_{a} =J_{a\sharp}~,
\end{align}
one finds
\begin{align}
[\hat J_{ab},\hat J_{cd}]&=
i(
\eta_{bc}\hat J_{ad}
-\eta_{ac}\hat J_{bd}
-\eta_{bd}\hat J_{ac}
+\eta_{ad}\hat J_{bc}
)~,
\nonumber\\
{[}\hat J_{ab}, \hat P_{c}]&=
i(
\eta_{bc}\hat P_{a}
-\eta_{ac}\hat P_{b}
)~,
~~~
{[}\hat P_{a},\hat P_{b}]=i\hat J_{ab}~.
\label{so(2,4)}
\end{align}
The $\hat J$ and  $\hat P$ are generators of the Lorentz  and translation symmetries
in the AdS$_5$ space.

The $R_{ij}$ ($i,j=4,5,6,7,8,9$) generates so(6).
The $\hat J_{a'b'}$ and $\hat P_{a'}$
($a',b'=5,6,7,8,9$)
defined by
\begin{align}
\hat J_{a'b'}=R_{ij}~,~~~
\hat P_{a'}=R_{4a' }~,
\end{align}
are generators of the Lorentz  and translation symmetries  in S$^5$
\begin{align}
[\hat J_{a'b'},\hat J_{c'd'}]&=
i(
\delta_{b'c'}\hat J_{a'd'}
-\delta_{a'c'}\hat J_{b'd'}
-\delta_{b'd'}\hat J_{a'c'}
+\delta_{a'd'}\hat J_{b'c'}
)~,
\nonumber\\
{[}\hat J_{a'b'}, \hat P_{c'}]&=
i(
\eta_{b'c'}\hat P_{a'}
-\eta_{a'c'}\hat P_{b'}
)
~,
~~~
{[}\hat P_{a'},\hat P_{b'}]=-i\hat J_{a'b'}~.
\label{so(6)}
\end{align}

Next we examine fermionic generators.
Define $\CQ_I$ ($I=1,2$) by
\begin{align}
\CQ_1\equiv Q+S\Gamma_4~,~~~
\CQ_2\equiv (Q-S\Gamma_4)\Gamma^{0123}~,
\label{fermionic: CFT-AdS}
\end{align}
and then one finds that   (anti-)commutators including $Q$ or $S$
are rewritten as
\begin{align}
[\CQ_I,\hat J_{ab}]&=-\frac{i}{2}\CQ_I \Gamma_{ab}~,~~
[\CQ_I,\hat J_{a'b'}]=-\frac{i}{2}\CQ_I \Gamma_{a'b'}~,~~
\nonumber\\
{[}\CQ_I,\hat P_a]&=\frac{i}{2}\epsilon_{IJ}\CQ_J\Gamma_a\CI~,~~
{[}\CQ_I,\hat P_{a'}]=-\frac{i}{2}\epsilon_{IJ}\CQ_J\Gamma_{a'}\CI~,~~
\nonumber\\
\{\CQ_I,\CQ_J\}&=
 \delta_{IJ} \left( C\Gamma^a \hat P_a
+C\Gamma^{a'}\hat P_{a'}
\right)h_+
-\frac{1}{2}\epsilon_{IJ} \left(
C\Gamma^{ab}\CI\hat J_{ab}
-C\Gamma^{a'b'}\CI\hat J_{a'b'}
\right)h_+~,
\label{super AdS5}
\end{align}
where we have defined $\CI\equiv \Gamma^{01234}$.
The superalgebra \bref{so(2,4)}, \bref{so(6)} and \bref{super AdS5}
is nothing but the supersymmetry algebra of the AdS$_5\times$S$^5$ background.
In fact, after the replacement $J_{AB}\to iJ_{AB}$, $P_A\to i P_A$
and $Q\to \frac{i}{\sqrt{2}}Q$,
the superalgebra \bref{so(2,4)}, \bref{so(6)} and \bref{super AdS5}
coincides with that in Ref.~\refcite{HKS02}
which gives the \adss{5}{5} superalgebra \cite{MT98}
in the ten-dimensional spinor notation.

Summarizing, we have shown that $\CN=4$ superconformal algebra
\bref{conf4} and \bref{super conf4}
is isomorphic
to the  superalgebra of AdS$_5\times$S$^5$,
\bref{so(2,4)}, \bref{so(6)} and \bref{super AdS5}.

Finally, we shall comment on the enhanced supersymmetries in the near-horizon limit
of the D3-brane solution
and the conformal supersymmetries of $\CN=4$ SYM.
The \adss{5}{5} background is the near horizon limit of the D3-brane solution.
The D3-brane solution preserves a half of 32 supersymmetries, 
while the \adss{5}{5} background does
full 32 supersymmetries.
In the near-horizon limit, 16 broken supersymmetries are restored 
so that the number of supersymmetries  enhances to 32.

For a D3-brane whose world-volume is extending along $0,1,2,3$ directions,
unbroken supersymmetry is specified by $\CQ_1+\CQ_2\Gamma^{0123}=0$,
while broken supersymmetry  by $\CQ_1-\CQ_2\Gamma^{0123}=0$.
Now we know the concrete expression \bref{fermionic: CFT-AdS} 
which relates fermionic symmetries in the  $\CN=4$ SYM and 
those in the \adss{5}{5} background.
We can see explicitly from \bref{fermionic: CFT-AdS} 
that
\begin{align}
\CQ_1+\CQ_2\Gamma^{0123}&=Q+S\Gamma^4+(Q-S\Gamma^4)(\Gamma^{0123})^2=2S\Gamma^4~,\\
\CQ_1-\CQ_2\Gamma^{0123}&=Q+S\Gamma^4-(Q-S\Gamma^4)(\Gamma^{0123})^2=2Q~.
\end{align}
This shows that the  unbroken supersymmetry on D3-brane
is related to the  supersymmetry $Q$ of the $\CN=4$ SYM,
while the enhanced supersymmetry in the near-horizon limit is
noting but the conformal supersymmetry $S$ of  the $\CN=4$ SYM.

\section{Summary and Discussion}

We have derived the $\CN=4$ superconformal algebra
from the symmetry transformations of fields in the  $\CN=4$ SYM in $D=4$.
As the $\CN=4$ SYM action in $D=4$ may be obtained from the $\CN=(1,0)$ SYM action
in  $D=10$ 
via a dimensional reduction,
the $\CN=4$ supersymmetry in $D=4$ can be expressed by using a MW spinor in $D=10$.
We found the map from generators of the $\CN=4$ superconformal algebra
to those of the super-\adss{5}{5} algebra.
Especially
we obtained the map  \bref{fermionic: CFT-AdS}
which relates fermionic generators:
the supersymmetries $Q$ and the conformal supersymmetries $S$
in $\CN=4$ SYM and
the supersymmetries $(\CQ_1,\CQ_2)$ in the \adss{5}{5} background.
It is obvious from this map that the conformal supersymmetries $S$ in  
the $\CN=4$ SYM 
correspond to the supersymmetries which are broken by a D3-brane and are restored in the near horizon limit.

In this note, we have examined the classical symmetry of the $\CN=4$ SYM.
An important generalization is to discuss the quantum symmetry,
such as the Ward-Takahashi identity.
For the $\CN=2$ SYM, 
the anti-commutators of Noether supercharges
are discussed in Ref.~\refcite{FO}
and central charges  are derived.
Applying this method to the $\CN=4$ SYM,
we find the following commutation relations
\begin{align}
\{Q_\alpha,
Q_\beta\}=&
\frac{1}{2}(C\Gamma^\mu h_+)_{\alpha\beta}P_\mu
+\frac{1}{2} Z_{\alpha\beta}~,\nn\\
\{S_\alpha,
S_\beta\}=&
\frac{1}{2} (C\Gamma^\mu h_-) _{\alpha\beta }K_\mu 
+\frac{1}{2} X_{\alpha\beta}~,
\nn\\
\{Q_\alpha,
S_\beta\}=&
\frac{1}{4} (C\Gamma^{\mu\nu} h_-)_{\alpha\beta} L_{\mu\nu}
-\frac{1}{2} (Ch_-)_{\alpha\beta}D
-\frac{1}{4} (C\Gamma^{ij} h_-)_{\alpha\beta} R_{ij}
+\frac{1}{2} Y_{\alpha\beta}~,
\end{align}
where $Z_{\alpha\beta}$, $X_{\alpha\beta}$ and $Y_{\alpha\beta}$
denote central charges
of the form
\begin{align}
Z_{\alpha\beta}=&
\int\!\d^3x \Big[
(C\Gamma^\mu h_+)_{\alpha\beta} 
A_\mu^a \Pi^{a0}
+(C\Gamma^i h_+)_{\alpha\beta}
\phi_i^a \Pi^{a0}
+\cdots
\Big]
~,
\nn\\
X_{\alpha\beta}=&
\int \!\d^3 x \Big[
(C\Gamma^\mu  h_-)_{\alpha\beta} (x^2\delta_\mu^\nu-2x_\mu x^\nu) \Pi^{a0}A_\nu^a
+(C\Gamma^i h_-)_{\alpha\beta} x^2 \Pi^{a0} \phi^a_i
+\cdots
\Big]
~,
\nn\\
Y_{\alpha\beta}=&
\int \!\d^3 x \Big[
-(C h_-)_{\alpha\beta} x^\mu A^a_\mu \Pi^{a0}
+ (C\Gamma^{\mu\nu} h_-)_{\alpha\beta}x_\mu A_\nu^a \Pi^{a0}
\nn\\&~~~~~~~~~
+(C\Gamma^{\rho i} h_-)_{\alpha\beta} x_\rho\phi^a_i \Pi^{a0}
+\cdots
\Big]
~.
\end{align}
The abbriviations  represent terms including fermions.
We have introduced $\Pi^{a\mu}$ defined by
$
\Pi^{a\mu}\equiv
D_\nu F^{a\nu\mu}-gf^{abc}\phi^b_i D^\mu \phi_i^c
+\frac{i}{2}gf^{abc}\bar\psi^b\Gamma^\mu \psi^c
$.
These central charges vanish if the equation of motion $\Pi^{a\mu}=0$
(and the fermionic counterpart) is used.
It is interesting to examine a realization of these central charges
in the AdS side
by using the map derived in this note.
We hope to report the complete form of these central charges and
discuss this point in the next work\,\cite{ST2}.

Another important generalization
is  to examine supersymmetries in AdS$_4$/CFT$_{3}$.
It is known that both  the $\CN=6$ superconformal algebra of
 the $\CN=6$ Chern-Simons-matter (CSM) theory 
(the ABJM model \cite{ABJM})
 and the super-isometry algebra of the AdS$_4\times\bbC$P$^3$ background
 are osp(6$|$4).
As in the AdS$_5$/CFT$_{4}$ case,
we can show that the conformal supersymmetries in the $\CN=6$ CSM
are supersymmetries which are broken by an M2-brane and are restored in the near horizon limit.
We will report this subject in near future.

\appendix

\section{Fierz identities in $D=10$}

The spin-dimension in $D=10$ is 32.
The following matrices composed of Gamma-matrices $\Gamma^M$ 
form a complete set of $32\times 32$ matrices
\begin{align}
\lambda^A=\Big\{&
1,\Gamma^M,\frac{i}{\sqrt{2!}}\Gamma^{MN},
\frac{i}{\sqrt{3!}}\Gamma^{M_1\cdots M_3},
\frac{1}{\sqrt{4!}}\Gamma^{M_1\cdots M_4},
\frac{1}{\sqrt{5!}}\Gamma^{M_1\cdots M_5},\nonumber\\&
\frac{i}{\sqrt{6!}}\Gamma^{M_1\cdots M_6},
\frac{i}{\sqrt{7!}}\Gamma^{M_1\cdots M_7},
\frac{1}{\sqrt{8!}}\Gamma^{M_1\cdots M_8},
\frac{1}{\sqrt{9!}}\Gamma^{M_1\cdots M_9},
\frac{i}{\sqrt{10!}}\Gamma^{M_1\cdots M_{10}}
\Big\}~.
\end{align}
The matrices $\lambda_A$ are defined by
$\mathrm{tr }\lambda^A\lambda_B = {32}\delta^A_B$.

Let $\phi_i$ ($i=1,2,3,4$) be Grassmann-even spinors in $D=10$,
and let  $\Lambda_1$ and $\Lambda_2$ be $32\times32$ matrices,
and then 
we have the identity\footnote{
The right hand side acquires an extra minus sign for Grassmann-odd spinors.}
\begin{align}
(\bar\phi_1\Lambda_1\phi_2)(\bar\phi_3\Lambda_2\phi_4)
=\frac{1}{32}\sum_{A}^{} (\bar\phi_1\lambda^A\phi_4)(\bar\phi_3\Lambda_2\lambda_A\Lambda_1\phi_2)~.
\end{align}

When $\phi_2=\phi_3=\phi_4\equiv \phi$
and $\phi_i$ have the same chirality $h_+\phi_i=\phi_i$,
we have
\begin{align}
(\bar\phi_1 \Gamma^M\phi)(\bar\phi \Gamma_M \phi)=0~.
\label{Fierz 1}
\end{align}
This is nothing but the Fierz identity \bref{10D Fierz}.

Finally we derive the Fierz identity \bref{Fierz SS}.
Let  $\epsilon$, $\psi$ and $\eta$ be Grassmann-odd spinors
of $h_+\epsilon=\epsilon,~h_+\psi=\psi$ and  $h_-\eta =\eta$.
For our purpose, we derive two Fierz identities,
\begin{align}
\epsilon~\bar\eta \psi&=-\frac{1}{16}\left[
\psi~\bar\eta\epsilon
-\frac{1}{2!}\Gamma^{MN}\psi~\bar\eta\Gamma_{MN}\epsilon
+\frac{1}{4!}\Gamma^{M_1\cdots M_4}\psi~\bar\eta\Gamma_{M_1\cdots M_4}\epsilon
\right]~,
\label{F1}
\end{align}
and
\begin{align}
\Gamma^M \eta~\bar\epsilon \Gamma_M  \psi&=
-\frac{1}{16}\bigg[
D \psi~\bar\epsilon \eta
-\frac{1}{2!}(D-4)\Gamma^{MN} \psi~\bar\epsilon \Gamma_{MN} \eta
\nn\\&~~~~~~~~~~~
+\frac{1}{4!}(D-8)\Gamma^{M_1\cdots M_4} \psi~\bar\epsilon \Gamma_{M_1\cdots M_4} \eta
\bigg]~,
\label{F2}
\end{align}
where $D=10$.
The Fierz identity \bref{Fierz SS}
can be obtained as $-i$\bref{F1}$ +\frac{i}{2}$\bref{F2}.

\section*{Acknowledgments}

The authors would like to thank Takanori Fujiwara, Yoshifumi Hyakutake,
Hyeonjoon Shin,
Haruya Suzuki and Kentaroh Yoshida
 for useful comments.


\begin{thebibliography}{9}

 \bibitem{gauge/gravity}
  G.~'t Hooft,
  Nucl.\ Phys.\ B {\bf 72} (1974) 461.

\bibitem{AdS/CFT}
J.~M.~Maldacena,
  Int.\ J.\ Theor.\ Phys.\  {\bf 38} (1999) 1113
   [Adv.\ Theor.\ Math.\ Phys.\  {\bf 2} (1998) 231]
  [hep-th/9711200].
  
  \bibitem{AdS/CFT2}
  S.~S.~Gubser, I.~R.~Klebanov and A.~M.~Polyakov,
  Phys.\ Lett.\ B {\bf 428} (1998) 105
  [hep-th/9802109].

E.~Witten,
  Adv.\ Theor.\ Math.\ Phys.\  {\bf 2} (1998) 253
  [hep-th/9802150].
 
  \bibitem{HKS02}
  M.~Hatsuda, K.~Kamimura and M.~Sakaguchi,
  Nucl.\ Phys.\ B {\bf 632} (2002) 114
  [hep-th/0202190].
  
\bibitem{MT98}
  R.~R.~Metsaev and A.~A.~Tseytlin,
  Nucl.\ Phys.\ B {\bf 533} (1998) 109
  [hep-th/9805028].

\bibitem{FO}
K.~Fujikawa and K.~Okuyama,
  Nucl.\ Phys.\ B {\bf 521} (1998) 401
  [hep-th/9708007].
  
    
\bibitem{ST2}
M.~Sakaguchi, F.~Takeuchi and K.~Yoshida,
work in progress.
  
\bibitem{ABJM}
  O.~Aharony, O.~Bergman, D.~L.~Jafferis and J.~Maldacena,
  JHEP {\bf 0810} (2008) 091
  [arXiv:0806.1218 [hep-th]].  


\end{thebibliography}
\end{document}